# Crystal lattice desolvation effects on the magnetic quantum tunneling of single-molecule magnets

G. Redler,[1] C. Lampropoulos,[2] S. Datta,[1] C. Koo,[1] T. C. Stamatatos,[2] N. E. Chakov,[2] G. Christou[2] and S. Hill[1,3,*]

[1]*Department of Physics, University of Florida, Gainesville, Florida 32611, USA*
[2]*Department of Chemistry, University of Florida, Gainesville, Florida 32611, USA*
[3]*Department of Physics and National High Magnetic Field Laboratory, Florida State University, Tallahassee, Florida 32310, USA*

**Abstract**
High-frequency electron paramagnetic resonance (HFEPR) and alternating current (AC) susceptibility measurements are reported for a new high-symmetry $Mn_{12}$ complex, $[Mn_{12}O_{12}(O_2CCH_3)_{16}(CH_3OH)_4]\cdot CH_3OH$. The results are compared with those of other high-symmetry spin $S = 10$ $Mn_{12}$ single-molecule magnets (SMMs), including the original acetate, $[Mn_{12}(O_2CCH_3)_{16}(H_2O)_4]\cdot 2CH_3CO_2H\cdot 4H_2O$, and the $[Mn_{12}O_{12}(O_2CCH_2Br)_{16}(H_2O)_4]\cdot 4CH_2Cl_2$ & $[Mn_{12}O_{12}(O_2CCH_2Bu^t)_{16}(CH_3OH)_4]\cdot CH_3OH$ complexes. These comparisons reveal important insights into the factors that influence the values of the effective barrier to magnetization reversal, $U_{eff}$, deduced on the basis of AC susceptibility measurements. In particular, we find that variations in $U_{eff}$ can be correlated with the degree of disorder in a crystal which can be controlled by desolvating (drying) samples. This highlights the importance of careful sample handling when making measurements on SMM crystals containing volatile lattice solvents. The HFEPR data additionally provide spectroscopic evidence suggesting that the relatively weak disorder induced by desolvation influences the quantum tunneling interactions, and that it is under-barrier tunneling that is responsible for a consistent reduction in $U_{eff}$ that is found upon drying samples. Meanwhile, the axial anisotropy deduced from HFEPR is found to be virtually identical for all four $Mn_{12}$ complexes, with no measurable reduction upon desolvation.

**I. Introduction**
Magnetic materials with nanoscale dimensions have attracted significant recent interest due to their potential application in emerging technologies [1], e.g. in high-density information storage devices and quantum computers. One possible route to nanoscale magnetic materials is the 'bottom-up' development of molecule-based magnets, or single-molecule magnets (SMMs [2,3]), which may one day supersede the 'top-down' approach involving nanoparticle-type materials found in most current nanoscale magnetic devices.

Molecular nanomagnets comprise a core of exchange-coupled transition metal ions (e.g., Mn, Fe, Ni or Co, etc.) that behave as single, rigid magnetic entities [4]. SMM properties arise when the (ferro- or ferri-) magnetic coupling between the metal ions results in a large ground state spin ($S$) combined with a significant and negative (easy-axis) magnetoanisotropy, as measured by the axial zero-field splitting (ZFS) parameter, $D$ [2,3]. This combination leads to a significant barrier $U$ to magnetization reversal whose maximum value is given approximately by $S^2|D|$. The barrier results in magnetic bistability and the ability to magnetize individual molecules below a characteristic blocking temperature, $T_B$ [5]. SMMs offer all of the advantages of molecular chemistry, while displaying the properties of much larger magnetic particles prepared by conventional top-down or miniaturization approaches. Most significantly, experimental studies of SMMs have facilitated important theoretical insights into a host of spectacular quantum phenomena, including magnetic quantum tunneling (MQT [6,7]) and quantum phase interference effects associated with tunneling magnetic particles [8] (see also [2,3]).

The molecular approach is particularly attractive due to the highly ordered and monodisperse nature of the molecules in the solid state, and because one may systematically vary many key parameters that influence the quantum behavior of a magnetic molecule, e.g., the total spin, symmetry, etc. [9,10,11]. Synthetic inorganic chemistry has provided a range of molecular clusters behaving as SMMs, with the $S = 10$ $Mn_{12}$ family of complexes being the most widely studied. The archetypal SMM, $[Mn_{12}O_{12}(O_2CCH_3)_{16}(H_2O)_4]\cdot 2CH_3CO_2H\cdot 4H_2O$ [hereon abbreviated $Mn_{12}Ac$ (**1**)], can be altered in a variety of ways in order to obtain new SMMs with modified or improved properties [12]. Approaches include substitution of the $CH_3CO_2^-$ carboxylate ligand with bulkier ones such as $BrCH_2CO_2^-$, $^tBuCH_2CO_2^-$, among others [11]; or exchange of the water molecules with alcohols [13]. These modifications primarily influence the local environment of the $[Mn_{12}O_{12}]^{16+}$ core, without affecting its total spin. In this way, one may affect subtle variations in the symmetry/anisotropy of the cluster. One can also reduce the $[Mn_{12}O_{12}]^{16+}$ core by adding electrons, with up to three electrons having been successfully added to date [11,14,15,16,17]; this approach may be used to realize half-integer spin states for studies of e.g. parity effects [18,19]. In this study, we focus on a sub-set of relatively high symmetry ($\sim S_4$) integer spin $S = 10$ complexes, while emphasizing the influence of the ligand/solvent environment on the quantum dynamics.

Only recently has a complete understanding emerged concerning the symmetry breaking responsible for the MQT in **1** [20,21,22,23,24,25,26]. While the average $S_4$ symmetry allows MQT interactions (transverse anisotropy) which are 4[th] order in the spin operators, intrinsic disorder associated with the non-centrosymmetric acetic acid solvent is found to lead to discrete local environments, resulting in a significant fraction of molecules possessing lower than four-fold symmetry [20-26]. Indeed, single-crystal high-frequency EPR (HFEPR) suggests that ~50% of the molecules experience a significant rhombicity, characterized by a 2[nd] order rhombic ZFS parameter $E \sim 0.01$ cm$^{-1}$ ($E/D \sim 0.02$) [21-26]. Soon after this realization, two ligand substituted high-symmetry $Mn_{12}$ variants were synthesized, $[Mn_{12}O_{12}(O_2CCH_2Br)_{16}(H_2O)_4]\cdot 4CH_2Cl_2$ ($Mn_{12}BrAc$ (**2**) [25,26,27,28]) and $[Mn_{12}O_{12}(O_2CCH_2Bu^t)_{16}(MeOH)_4]\cdot MeOH$ ($Mn_{12}Bu^tAc$ (**3**) [25,26,29]). Effects attributed to the disordered acetic acid in **1** were found to be absent in **2** and **3**, thus revealing the quantum dynamics of $Mn_{12}$ in a truly axially symmetric ($S_4$) environment for the first time [25-30].

Here, we report on a new $Mn_{12}$ complex that is still more closely related to the original acetate: $[Mn_{12}O_{12}(O_2CCH_3)_{16}(CH_3OH)_4]\cdot CH_3OH$ [$Mn_{12}Ac/CH_3OH$ (**4**)]. We also explore further the effects of disorder on the quantum dynamics of $Mn_{12}$ SMMs by deliberately removing solvent from crystals of the new high-symmetry complexes (**2–4**) by drying them under vacuum. In particular, we establish a clear correlation between the degree of disorder and the magnitude of the effective barrier, $U_{eff}$, to magnetization relaxation deduced from alternating current (AC) susceptibility measurements. We also find from HFEPR studies that solvent loss/disorder has very little effect on the axial ZFS parameter, $D$, while inducing significant transverse anisotropy. We thus conclude that the disorder-induced reduction in $U_{eff}$ is caused by enhanced under-barrier

MQT, not by a reduction in *D*. These findings may have important implications for previous studies of SMMs [31,32,33,34].

## II. Experimental details

The structure of complex **4**, shown in Fig. 1, is very similar to the other high-symmetry $Mn_{12}$ complexes considered in this study [28,29,35]. Nevertheless, it does exhibit several important differences compared to its closest relative—$Mn_{12}$Ac (**1**): (i) the four terminal water molecules are replaced by terminal methanols ($CH_3OH$); (ii) the two acetic acid and four water solvent molecules are replaced by only one methanol; and (iii) this methanol solvent molecule resides on a symmetry element, making the overall structure highly symmetric (crystallographic space group $I\bar{4}$). A direct consequence of (iii) is the retention of overall molecular $S_4$ symmetry (including the solvent of crystallization). Therefore, this version of $Mn_{12}$ represents a much cleaner, high-symmetry derivative of the original $Mn_{12}$Ac, much like the recently reported complexes **2** and **3**. The synthesis of complex **4** will be reported elsewhere.

In order to measure the AC magnetic susceptibility of pristine (wet) samples of **4**, the crystalline material was first removed from the mother liquor and rapidly transferred to an analytical balance for accurate weight measurement. Then, within one minute, the crystals were carefully embedded in eicosane within a gelatin capsule in order to ensure retention of the solvent of crystallization. Dry samples of **4** were obtained by first removing crystals from their mother liquor, then drying under vacuum for a period of ~6 hours. These samples were then carefully weighed and embedded in eicosane as described above. Measurements of the in-phase ($\chi'$) and out-of-phase ($\chi''$) AC susceptibility were made in the frequency range from 5 Hz to 1.5 kHz using a Quantum Design MPMS system. A similar procedure was followed for complexes **1** − **3**. A fresh batch of **1** was prepared for this investigation (see also earlier reports [36,37,38]); the results for complexes **2** and **3** are reported elsewhere ([28] and [29], respectively).

HFEPR measurements were performed on single-crystals of **4** at various frequencies in the range from 50 to 400 GHz using a high-sensitivity cavity perturbation technique and a Millimeter-wave Vector Network Analyzer (MVNA) described in elsewhere [39,40]. The magnetic field was provided by a 7 T horizontal-bore superconducting magnet associated with a Quantum Design PPMS system. The horizontal-bore magnet facilitates in-situ rotation of the cavity relative to the applied field. The cavity additionally permits further rotation of the sample about an orthogonal axis (in-situ), thereby enabling collection of data for both easy-axis ($B//c$-axis) and hard-plane ($B \perp c$-axis) orientations. Sample alignment is first achieved by locating extrema among plots of spectra recorded at many different field orientations; once aligned, multi-frequency measurements are performed in order to provide data sets which maximally constrain the ZFS parameters. Similar procedures were followed for complexes **1** − **3** which are reported elsewhere [21-29,41].

The typical crystal size (needles with length ~200μm and diameter ~30μm) proved marginally too small to obtain sufficient quality spectra at the high frequencies (>300 GHz) required for constraining the axial ZFS parameters (*D* and $B_4^0$—see below); note that the sensitivity of our spectrometer diminishes considerably for frequencies above 200 GHz, partly due to reduced signal-to-noise and partly because the cavity is highly over-moded at these high frequencies. We therefore carefully positioned ~20 aligned needle shaped single-crystals on a quartz pillar coincident with the axis of the cylindrical cavity. This procedure ensured alignment of the easy-axes to within a few degrees (< ±10°), thereby enabling reliable analysis of the easy-axis spectra due to the relative insensitivity of the peak positions ($\propto \cos\theta$) for fields close to this orientation. In fact, it also proved possible to obtain reasonable hard-plane spectra (not shown) in spite of the mosaicity of the polycrystalline sample. Finally, we note that the crystals of complex **4** used for HFEPR studies were quickly transferred from the mother liquor and coated in silicone grease in order to avoid solvent loss. The samples were also initially cooled under atmospheric helium gas, with a total transfer time from the mothor liquor to the cryostat of just 15-20 minutes. All of the HFEPR data for complexes **1 – 3** presented here have been reported previously [22-28].

### III. Results
### III(a) AC susceptibility

Figures 2(a) and (b) respectively display the low-temperature behavior of $\chi'$ and $\chi''$ for complex **4** (wet) for eight different frequencies in the range from 5 Hz to 1.5 kHz. For each frequency, $f$, the sharp drop in $\chi'$ and the corresponding peak in $\chi''$ marks the onset of blocking of the magnetization. An Arrhenius plot of $\ln(2\pi f)$ versus $1/T_M$ is displayed in Fig. 3, where $T_M$ corresponds to the temperature at which the maximum in $\chi''$ is observed. The values of $T_M$ were obtained by fitting a Lorentzian function to 3 or 4 data points either side of the presumed maxima in Fig. 2(b). Even the most conservative estimates of the uncertainties associated with this procedure result in error bars that are smaller than the data symbols ($< 4 \times 10^{-4}$ K$^{-1}$) employed in Fig. 3.

The behavior observed in Figs. 2 and 3 is very typical for a good SMM, and the procedure outlined here is the method of choice for determining the effective barrier height ($U_{eff}$) to magnetization reversal from a linear fit to an Arrhenius plot of the form

$$\ln(1/\tau) = \ln(1/\tau_o) - U_{eff}/k_B T_M; \qquad (1)$$

here, $\tau$ (= $1/2\pi f$) is the relaxation time and $\tau_o$ the attempt time for reversal at $T = \infty$ [28]. Of course, this procedure makes a number of assumptions—not least the notion that there exists just a single relaxation pathway over a well defined (classical) anisotropy barrier, i.e. $(1/\tau) = (1/\tau_o) \exp\{-U_{eff}/k_B T_M\}$. Nevertheless, the values of $U_{eff}$ and $T_B$ represent the primary benchmarks used by synthetic inorganic chemists to compare and, ultimately, rank SMMs (the latter is determined from DC coercivity measurements and generally scales with $U_{eff}$).

A subsidiary message of the present investigation is that one must exercise extreme caution when making comparisons or basing conclusions upon relatively small (~10-15%) variations in $U_{eff}$ (or, for that matter, $\tau_o$). Before worrying about sample handling and effects caused by disorder and/or solvent loss, we note the very obvious deviation of the data from an Arrhenius law in the main panel of Fig. 3, i.e. there is a definite curvature, leading to an apparent increase in slope ($U_{eff}$) at higher frequencies; for clarity, data for the wet sample have been emphasized in the main panel (black line and black squares). This behavior, which has been noted by previous authors [37,38], seems to be reproducible among high-symmetry Mn$_{12}$ complexes. It is an obvious indication for a deviation from simple Debye relaxation [42], leading to a situation in which published values of $U_{eff}$ are often significantly larger (by >5%) than the barrier associated with the giant spin ground state ($S = 10$ for high symmetry Mn$_{12}$) estimated on the basis of spectroscopic measurements such as HFEPR [43] or inelastic neutron scattering (INS [44,45]).

Furthermore, one can see that the actual values of $U_{eff}$ obtained from fits to an Arrhenius law [Eq. (1)] will depend significantly on the range of frequencies chosen for AC susceptibility measurements. For this reason, one can often find significant variations in $U_{eff}$ values throughout the literature [36-38]. As we shall show here, part of the reason is due to the non-Arrhenius behavior observed in Fig. 3. However, another important factor involves sample handling and preparation due to the possibility of solvent loss, which is the main focus of this stusy.

No attempt is made here to account for the deviation from Arrhenius behavior (see [42]). However, in making comparisons between $U_{eff}$ values for various samples, we have made sure to employ AC data sets recorded at *exactly* the same eight (or four) frequencies as those displayed in Figs. 2 and 3. Table 1 summarizes the values of $U_{eff}$ obtained for wet and dry samples of complexes **1 – 4** based on eight- and four-point fits to Eq. (1) [46]: the former from fits to the full data sets (see main panel of Fig. 3) and the latter from fits to the lowest four frequencies (see inset to Fig. 3). The $\tau_o$ values obtained from the eight-point fits are also tabulated, although we comment only briefly on them in this work. There are several points to note about the $U_{eff}$ values displayed in Table 1: (1) because of the non-Arrhenius behavior, the effective barriers deduced on the basis of the eight-point fits exceed those from the four-point fits by ~3-5 K; (2) the values of $U_{eff}$ obtained for the wet samples from the four-point fits are in broader agreement with the barriers obtained from EPR experiments ($U_{EPR}$ ~ 66−68 K, *vide infra*–see Table 2); (3) the uncertainties associated with the $U_{eff}$ values obtained from the four point-fits are significantly less than those associated with the eight-point fits; and (4) the effective barriers are depressed by between 2 and 7 K upon drying the crystals, with the effect being by far the most pronounced for complex **2**.

It is important to recognize that the values of $U_{eff}$ given in Table 1 contain significant systematic errors associated with the fact that the data do not follow an Arrhenius law. The numerical uncertainties given in the table do not fully reflect this error. As can clearly be seen, one obtains a significant variation in $U_{eff}$ upon sampling different ranges of the data; indeed, this variation (3−5 K) far exceeds the uncertainties given in Table 1 (< 1 K). However, the Arrhenius approximation improves as $T \to 0$. For this reason, the fits improve when sampling only the lowest four temperatures/frequencies, as reflected in the corresponding uncertainties given in Table 1. This enables comparisons between data sets obtained at exactly the same four frequencies (and approximately the same four temperatures). In other words, on this basis, the ~2 K reduction in $U_{eff}$ deduced from the four point fits (uncertainties ~0.5 K) is statistically significant, as is also clearly evident simply from a visual inspection of the data displayed in the inset to Fig. 3.

One may now draw several conclusions from the data in Table 1. As noted above, the first and most obvious conclusion is the clear trend involving the effective barrier reduction upon drying samples of complexes **2 – 4** (see also Fig. 3). Attempts to dry **1** at ambient temperatures were not successful, most probably due to the strongly hydrogen bonding nature of the solvents. While the drying effect for complexes **3** and **4** is small (~2 K reduction in $U_{eff}$), it is statistically significant. Meanwhile, the effect observed for complex **2** is quite dramatic (a reduction of ~7-8 K). Of the various solvents found in the lattices of complexes **1 – 4**, $CH_2Cl_2$ is by far the most volatile (boiling point is ~40ºC), thus likely explaining the dramatic effect upon drying **2**. Indeed, crystals of **2** observed under a microscope are seen to dry and crack when left in air for just a few minutes. In contrast, samples of **1** are known to retain their properties (and luster) for many months (even years) after removal from their mother liquor. Meanwhile, **3** and **4** have been shown to lose methanol solvent upon drying [29]. Therefore, we find a clear correlation between solvent volatility and effective barrier reduction upon drying samples.

Another important conclusion concerns the magnitudes of the obtained effective barriers. As noted above, $U_{eff}$ is simply an adjustable parameter in a less than perfect model [42]. To a certain degree, this is also the case with the barrier deduced from EPR ($U_{EPR}$) on the basis of fits to a giant spin Hamiltonian (GSH), and there is no reason to expect agreement between these numbers. However, most properties of $Mn_{12}$ can be explained quite well using a GSH [21-30]. We therefore have good reason to presume that the barrier heights obtained from HFEPR (and INS) should be quite reliable; indeed, the experimental uncertainties given in Table 2 truly reflect the confidence in the corresponding values of $U_{EPR}$ (for further discussion, see following section). The values of $U_{eff}$ obtained from AC susceptibility data are in-fact more likely to be lower than the barriers ($U_{EPR}$) obtained from HFEPR analysis based on the GSH, because MQT often short-circuits the quantum states near the very tops of the barriers [47,48]. It is thus clear that the higher frequency AC data contribute to the over-estimation of $U_{eff}$. An explanation for this behavior lies beyond the scope of the present study which focuses on the influence of solvent loss and disorder on the relaxation dynamics of high-symmetry $Mn_{12}$. We nevertheless conclude that the lower frequency AC data (i.e. the four-point fits) provide a more reliable benchmark for comparison with spectroscopic data (e.g. HFEPR) [42], and that Eq. (1) correctly captures the spin dynamics associated with the (Debye) relaxation of a rigid $S = 10$ object over a single barrier in the limit $T \rightarrow 0$, deviating from this simple picture at elevated temperatures/frequencies. For this reason, we note that values quoted in the literature often overestimate $U_{eff}$ by up to 15% [43] (in some cases by as much as a factor of 2.5 [38]!!).

For the remainder of the paper, we focus on obtaining an understanding of the effective barrier reduction induced by sample drying via comparisons between HFEPR studies and the four-point fits given in Table 1. We emphasize that by using the same four frequencies, one can make meaningful comparisons between various samples, particularly wet and dried versions of the same compound. Although drying clearly introduces significant uncertainties in the absolute values of $\chi'$ and $\chi''$, it does not influence our systematic approach for estimating $U_{eff}$, as described above. Certainly, the comprehensive study presented here, involving several different $Mn_{12}$ complexes, paints a consistent picture in terms of the effect of sample handling on $U_{eff}$.

## III(b) High Frequency EPR

Figure 4 displays a compilation of the EPR peak positions observed during magnetic field sweeps at different frequencies, with the field aligned parallel to the average easy- ($z$-) axes of the crystals in the sample. Superimposed upon the data is the best simulation obtained via solution of the GSH [3]

$$\hat{H} = D\hat{S}_z^2 + B_4^0 \hat{O}_4^0 + g_z \mu_B B \hat{S}_z, \qquad (2)$$

followed by an evaluation of the energy differences between eigenstates which are connected by the magnetic dipole transition operator, i.e. states that differ in spin projection, $m_s$, by ±1. Note that Eq. (2) contains only the total spin operator $\hat{S}$ (part of the $\hat{O}_4^0$ operator, defined elsewhere [3]) and powers of its $\hat{S}_z$ component; thus, all terms commute, i.e. this approximation assumes that the total spin, $S$, and its projection, $m_s$, onto the molecular easy axis are exact quantum numbers. The first and second terms in Eq. (2) respectively characterize the 2nd and 4th order axial anisotropies, parameterized by $D$ and $B_4^0$, and the values used for the simulation in Fig. 4 are listed in Table 2. The final term in Eq. (2) represents the Zeeman interaction, assuming the magnetic field, $B$, is parallel to $z$. The obtained Landé factor, $g_z = 1.92$, is a little on the low side, reflecting the fact that the crystals in the sample are not perfectly aligned. Therefore, the polycrystalline nature of the sample does not permit an accurate determination of the Landé factor. However, extrapolation of the data to zero-field by means of simulation provides very precise estimates of the zero-field splittings. Furthermore, the influence of anticipated transverse ZFS parameters [not included in Eq. (2)] is immeasurable at these high frequencies [24,49]. Consequently, the procedure outlined here provides highly reliable estimates of the axial anisotropy for **4**, and it is these parameters that directly determine $U_{EPR}$—the theoretical barrier height determined from EPR [49]. Values of $U_{EPR}$ estimated for complexes **1** – **4** (wet) are listed in Table 2.

Inspection of Table 2, indicates that the values of $U_{EPR}$ are almost the same for the four complexes: $U_{EPR}$ is slightly larger for complex **2** and marginally lower for complex **1**, as compared to **3** and **4**; however, these differences are barely statistically significant, and are also far smaller than the corresponding variations in $U_{eff}$. The small differences in $U_{EPR}$ are, perhaps, not surprising given that the $Mn_{12}O_{12}$ cores are essentially identical for the four complexes. This suggests that the scatter in the values of $U_{eff}$ is not due to variations in the intrinsic axial anisotropy ($D$ and $B_4^0$) associated with each cluster. Most likely, this reflects differences in the rates of under-barrier tunneling (*vide infra*) due to transverse ZFS interactions, either of intrinsic or extrinsic (disorder [21]) origin.

The above conclusions are further supported by comparisons between HFEPR data for wet and dry samples of **1** and **2** presented in Fig. 5 [23,28]; all data were collected on single crystal samples at 51.5±0.1 GHz and 15 K, with the magnetic field applied perpendicular to the easy axis of each crystal (in between the two orthogonal two-fold hard axes in the case of **1**—see ref. [23,24]). Note that the wet sample of complex **2** exhibits very sharp, symmetric EPR transitions with a significant peak-to-peak height variation; for an explanation of the labeling, see ref. [23]. The dry samples (A and B) were prepared in-situ by thermally cycling the crystal to ~290 K under vacuum; runs A and B were simply obtained after successive thermal cycling. In contrast to the wet case, the HFEPR spectra obtained from dried crystals of **2** exhibit significant line broadening and asymmetry, particularly α8 and α6, i.e. the effect is more pronounced at higher fields. This line broadening leads to a far more gradual peak-to-peak height variation (the peak-to-peak variation in the areas under the resonances is the same). A key observation is the fact that the spectra for dried samples of **2** end up looking exactly like the intrinsically asymmetric HFEPR spectra obtained for wet samples of **1** (see Fig. 5 and discussion below). A second important observation is the fact that the field locations of absorption maxima (transmission minima) do not shift significantly upon drying, though a slight shift to lower fields of the central moment of each transition is discernible.

If one were to assume that drying complex **2** results in a ~10% decrease in axial anisotropy, as inferred from the 10% decrease in $U_{eff}$ (Table 1), then the peaks observed in the HFEPR spectra in Fig. 5 should shift dramatically to lower fields, as indicated by the arrows and vertical dashed lines. Meanwhile, the relatively weak variation in the spectra observed for runs A and B suggests that the sample was already well dried after the first thermal cycling, yet no significant peak position shifts are observed. This provides dramatic proof that solvent loss has almost no measurable effect on the dominant axial cluster anisotropy ($D$ and $B_4^0$), which is perhaps not surprising given that the solvents generate only minor perturbations to the $Mn_{12}$ cores. In fact, the effects of solvent disorder are not completely insignificant, as has been studied in great detail for complex **1**, where intrinsic disorder is unavoidable [20-26]. The disorder-induced strains produce a roughly 2% full-width at half maximum for the distribution in the axial parameter $D$ [50-52]. More significant, however, is the effect of the disorder on the off-diagonal (transverse) anisotropy. In the absence of disorder, the cluster symmetry is high ($S_4$) and the 2$^{nd}$ order transverse ZFS term is strictly zero. For this reason, relatively weak disorder can give rise to appreciable off-diagonal ZFS, i.e. non-zero 2$^{nd}$ order rhombicity ($|E| \sim 0.01$ K). Indeed, simulations of the hard-plane HFEPR spectra obtained for wet crystals of **1** reproduce the observed asymmetric line shapes (see Fig. 5) only when known (statistical) disorder is taken into consideration [23].

Based on the observations outlined above, we take the view that wet crystals of **2** have an intrinsically high symmetry structure, i.e. there is little or no disorder in the crystal, and the $Mn_{12}$ cluster anisotropies are extremely monodisperse (this applies to *all* components of the ZFS tensor). Upon drying crystals of **2**, solvent is lost (not necessarily all), resulting in a weakly disordered lattice which more closely mirrors the situation found in wet crystals of **1** which suffer from a significant distribution in cluster anisotropies caused by an intrinsic disorder associated with the acetic acid solvent [20]. These findings are supported by magnetic studies of dried crystals of **2**, which find significant disorder [34]. In contrast, HFEPR studies of wet crystals of **2** reveal a far more ordered structure in comparison to wet crystals of **1** [25-28].

HFEPR studies of **1** have provided considerable insights into the nature of the disorder that gives rise to the asymmetric hard-plane EPR line shapes seen in Fig. 5 [23-26]. The asymmetry is due primarily to weak tilting of the local ZFS tensors (< 1°) combined with local 2$^{nd}$ order anisotropy that is otherwise forbidden. Meanwhile, the influence of the disorder on the axial terms is minimal [50,52]. Subsequent magnetic studies have demonstrated that the disorder-induced 2$^{nd}$ order anisotropy significantly influences the MQT observed in **1** [24].

## **IV. Further Discussion**

We are now in a position to put together the findings of a very broad range of studies of several closely related high-symmetry $Mn_{12}$ complexes. It is clear that the axial anisotropies associated with **1 – 4** are essentially identical, even after drying. Nevertheless, the act of drying clearly results in differing degrees of solvent loss (most pronounced for **2**), leading to weak disorder and to a reduction of $U_{eff}$. This effective barrier reduction appears to be driven by disorder-induced off-diagonal (transverse) ZFS interactions, i.e. the disorder increases tunneling, which is most pronounced near the top of the barrier, as illustrated in Fig. 6. The tunneling results in relaxation channels that short-circuit the very top of the classical barrier ($U_{EPR}$), which remains fixed at ~68 K. Consequently, relaxation studies perceive a reduced effective barrier, $U_{eff}$. While these ideas are not new, such clear cut experimental evidence for this phenomenology is unprecedented. It is particularly notable that the $U_{eff}$ value in Table 2 for the intrinsically disordered complex **1** is significantly lower than those of the other three complexes, particularly as compared to **2** and **3**. Meanwhile, the four-point $U_{eff}$ values in Table 1 for dried complexes **2** and **3** (also **4** to some extent) are essentially identical to those of wet crystals of **1**. We thus conclude that wet crystals of the newer $Mn_{12}$ complexes (**2** to **4**) are intrinsically ordered, and that drying converts them to a less ordered state which is similar to the original $Mn_{12}Ac$ (**1**).

This work once again highlights the critical role of lattice solvents and sample handling in experimental studies of SMMs: *the solvent molecules are not innocent!* In most instances, the low temperature ($T < T_B$) quantum dynamics of a SMM is governed by near immeasurable symmetry breaking terms which produce tunnel splittings that can be ten or more orders of magnitude (!!) smaller than the dominant energy scale, $U_{EPR}$. Therefore, it requires only very minor perturbations to completely alter the energy balance, thus dramatically influencing the low-temperature quantum dynamics. While the EPR data reported here were mostly collected at temperatures above the blocked regime ($T > T_B$), the high resolution of this spectroscopic technique is more than sufficient for characterizing these weak tunneling interactions, as we have demonstrated here and in many previous works [23-28]. A broad survey clearly demonstrates that crystals which form with no solvent in their lattice exhibit remarkably clean HFEPR spectra and MQT behavior [53,54,55,56]. Furthermore, such samples are amenable to thermal cycling without degradation. This is not the case for the vast majority of SMMs for which the solvent not only influences the line widths associated with EPR spectra (and MQT resonances) but also the quantum dynamics. We stress—it is not simply the quality of spectra that degrade as a result of solvent loss, the physics changes as well! For this reason, measurements that expose samples to conditions that favor solvent loss will likely yield different information concerning the low-temperature relaxation dynamics, as compared to studies of wet single crystals. This includes studies on powders or polycrystals pressed into pellets [57], as well as samples synthesized under extreme conditions, e.g. nanocrystals prepared in compressed fluids [31,33].

Finally, we have so far commented very little on the role of the pre-exponential factor $\tau_o$ [see Eq. (1)]. Once again, a clear trend is found from the data in Table. 1, namely that $\tau_o$ exhibits a roughly 50% increase upon drying (the errors associated with the values in the table are ~20%). This is likely related to sample crystallinity which clearly degrades upon drying; $\tau_o$ is ultimately related to the coupling of spins to lattice vibrations. Other recent studies have demonstrated a sensitivity of $\tau_o$ to sample size/morphology [32,33].

### V. Summary and Conclusions

We have performed HFEPR measurements on a new high-symmetry $Mn_{12}$ SMM (**4**), which differs from the original $Mn_{12}Ac$ (**1** [3,5,35]) both in terms of identity of the terminal ligands which, in the case of **4**, are methanols instead of waters, and in terms of the lattice solvents (one methanol in **4** vs. two acetic acids and four waters in **1**). Complex **4** has a spin $S = 10$ ground state and a barrier to magnetization reversal determined by HFEPR [$U_{EPR} = 67.4(13)$ K] which is essentially identical to those found for the other $Mn_{12}$ complexes considered in this study (**1** – **3**).

We compare values of $U_{EPR}$ with the effective barriers ($U_{eff}$) deduced via AC susceptibility measurements for all four complexes. In doing so, we consistently observe non-Arrhenius relaxation behavior which leads to significant problems in reliably estimating $U_{eff}$. We also observe a clear dependence of the low-temperature relaxation dynamics on whether a sample is wet or dry, which we parameterize in terms of an effective barrier deduced on the basis of a $T \to 0$ extrapolation. Differences between wet and dry samples are attributed to disorder in the latter, caused by the loss of lattice solvent. The combined HFEPR and magnetic measurements indicate that the faster relaxation found in the dried samples is caused by disorder-induced tunneling below the top of the classical barrier, $U_{EPR}$. This tunneling leads to an apparent reduction in the effective barrier, $U_{eff}$, to magnetization relaxation; the classical barrier ($U_{EPR}$), meanwhile, is relatively insensitive to the solvent disorder induced through drying.

**VI. Acknowledgements**

This work was supported by the NSF (CHE0414555, DMR0804408, DMR0239481 and DMR0506946). GR acknowledges the support of the University of Florida Undergraduate Scholars Program.

**VII. References Cited**

* Electronic address: `shill@magnet.fsu.edu`

**Table 1:** Comparison of wet vs. dry AC susceptibility data for complexes **1** to **4**.

| Complex | Solvent | Space group | $U_{eff}$ (K) 8 point | $U_{eff}$ (K) 4 point | $\tau_o$ (s) | Ref. |
|---|---|---|---|---|---|---|
| **1** (wet) | 2CH$_3$COOH, 4H$_2$O | $I\bar{4}$ [(1)] | 69.8(9) | 65.0(4) | 1.1 × 10$^{-8}$ | |
| **2** (wet) | 4CH$_2$Cl$_2$ | $I4_1/a$ | 76.8(8) | 72.5(5) | 2.0 × 10$^{-9}$ | [28] |
| **2** (dry) | | | 69.1(8) | 65.2(4) | 3.8 × 10$^{-9}$ | [28] |
| **3** (wet) | CH$_3$OH | $I\bar{4}$ | 71.0(9) | 66.2(9) | 5.3 × 10$^{-9}$ | [29] |
| **3** (dry) | | | 67.5(6) | 64.3(2) | 7.7 × 10$^{-9}$ | [29] |
| **4** (wet) | CH$_3$OH | $I\bar{4}$ | 74.4(9) | 69.2(5) | 5.7 × 10$^{-9}$ | |
| **4** (dry) | | | 71.9(9) | 67.1(6) | 9.8 × 10$^{-9}$ | |

$\tau_o$ = pre-exponential factor in the Arrhenius law obtained from 8 point fits.
[(1)] Crystallographic average symmetry [20].

**Table 2:** Comparison of $U_{EPR}$ and $U_{eff}$ for complexes **1** to **4** (wet).

| Complex | $D$ (K) | $B_4^0$ (K) | $U_{EPR}$ (K) | $U_{eff}$ (K)[1] | Ref. |
|---|---|---|---|---|---|
| 1 | −0.655(5) | −2.9(3) ×10$^{-5}$ | 66.2(13) | 65.0(4) | [23] |
| 2 | −0.674(3) | −3.6(2) ×10$^{-5}$ | 68.2(5) | 72.5(5) | [28] |
| 3 | −0.665(5) | −3.6(3) ×10$^{-5}$ | 67.3(12) | 66.2(9) | [25,26] |
| 4 | −0.667(5) | −3.1(3) ×10$^{-5}$ | 67.4(13) | 69.2(5) | |

[1] Four-point fit values for the wet samples from Table 1.

**Figure captions**

**Fig. 1:** (Color online) The structure of complex **4**. Color code: Mn$^{IV}$ green, Mn$^{III}$ blue, O red, C grey.

**Fig. 2:** (Color online) Temperature-dependent plots of (a) the real ($\chi'$) and (b) imaginary ($\chi''$) components of the AC magnetic susceptibility vs. temperature in a 3.5 G field, oscillating at the indicated frequencies.

**Fig. 3:** Arrhenius plots of maxima ($T_M$) observed in $\chi''$ for wet and dry crystals of complex **4** (see Fig. 2). Main panel: a linear fit to Eq. (1) for the full data set of eight frequencies for the wet sample; note the non-linearity of the points, which do not fit well to the linear regression. Inset: linear fits to the lowest four frequencies for both the wet and dry samples; the difference in slopes is a clear indication of the lower effective barrier associated with the dry sample.

**Fig. 4:** Frequency dependence of HFEPR peak positions (black squares) obtained with the field applied parallel to the average easy-axes of multiple aligned crystals of **4**; the temperature was 15 K for these measurements. Superimposed on the data is the best simulation based on the GSH given in Eq. (2); the obtained ZFS parameters are given in Table 2.

**Fig. 5:** (Color online) Comparison between hard plane EPR spectra obtained at 51.4±0.1 GHz and 15 K for complexes **1** and **2**. The data for **2** were obtained for wet and dry samples, and have been normalized according to the α2 intensity (see [23] for explanation of labeling); runs A and B were obtained after successive cycling of the same sample to room temperature. The data for **1** were obtained with the magnetic field applied in between the (two-fold) hard-axes associated with the low-symmetry disordered species (see [23,24] for further explanation). The fact that the peaks associated with **1** are observed at slightly lower fields than the peaks for **2** is a direct confirmation of the slightly larger *D* value for complex **2** (see Table 2).

**Fig. 6:** (Color online) Schematic illustrating the classical and quantum relaxation processes responsible for the observed effective barrier reduction upon drying samples—see main text for further explanation.

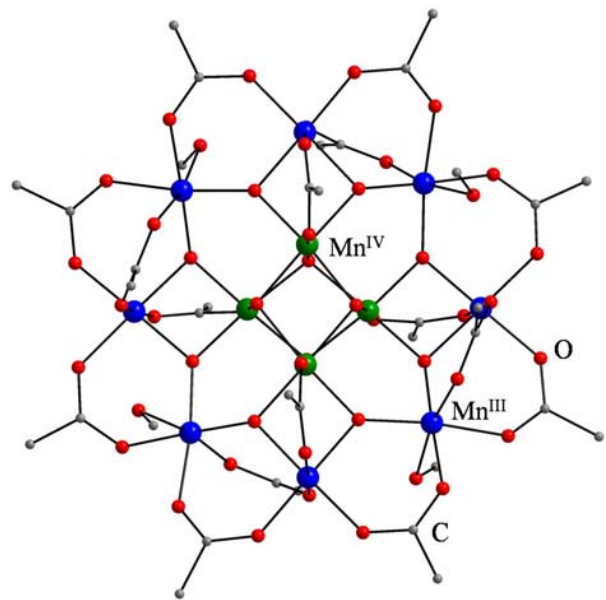

Figure 1, Redler et al.

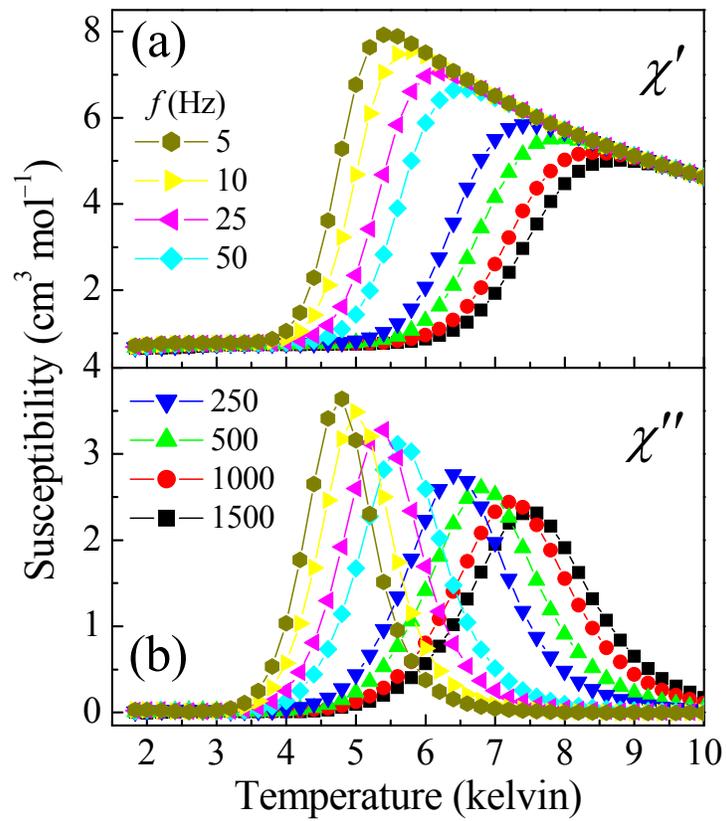

Figure 2, Redler et al.

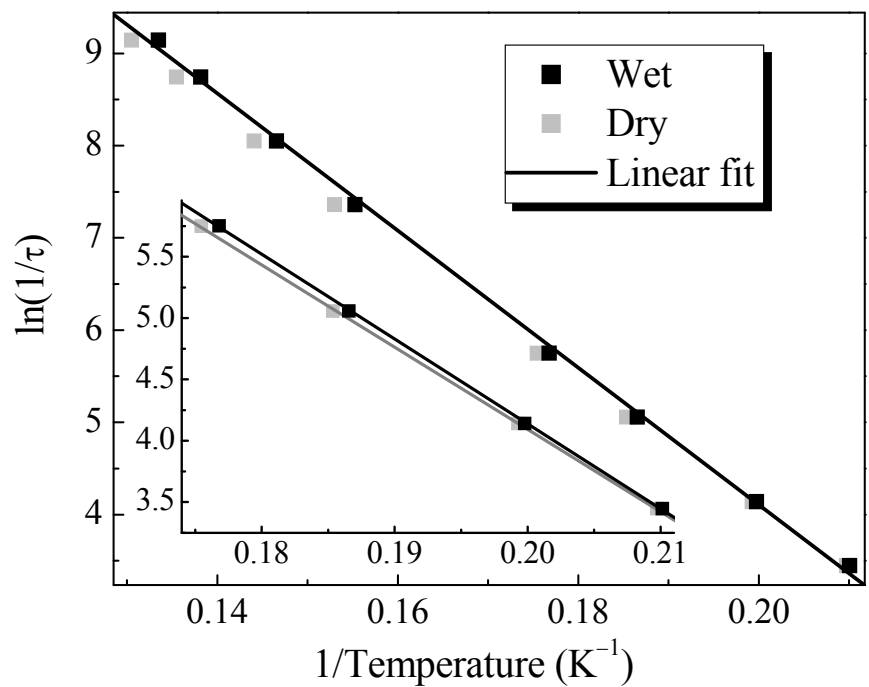

Figure 3, Redler et al.

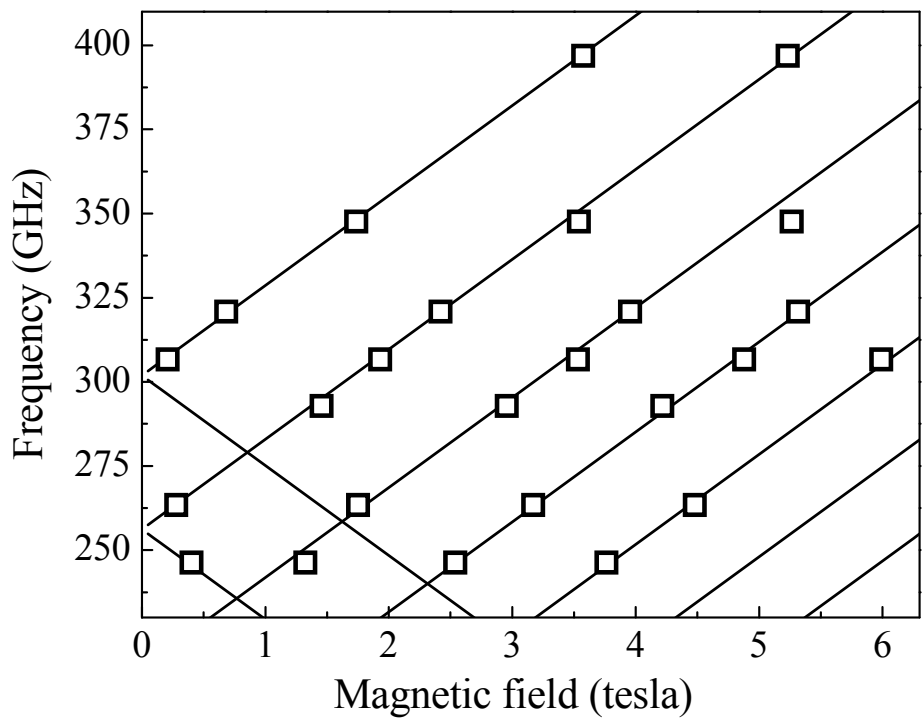

Figure 4, Redler et al.

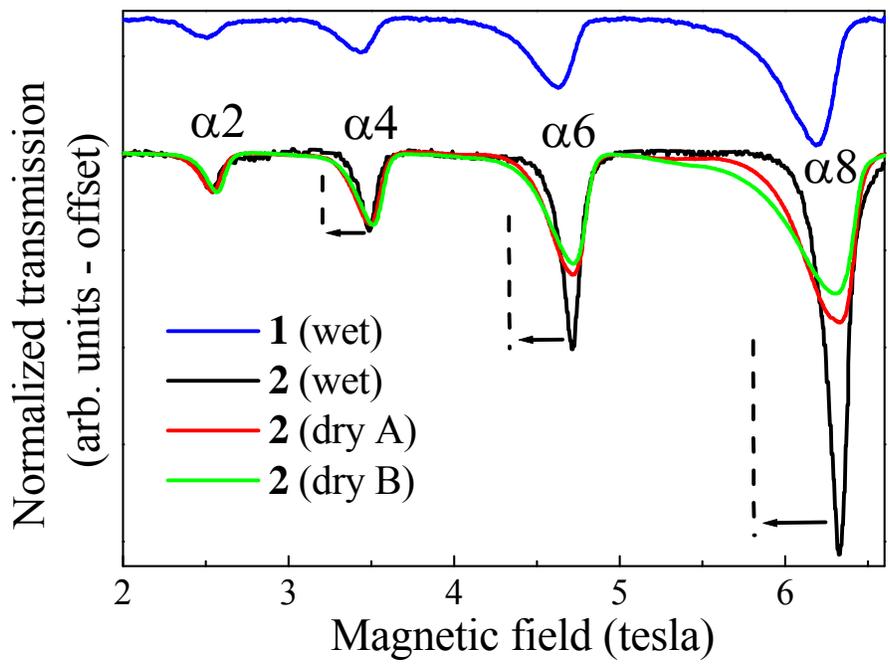

Figure 5, Redler et al.

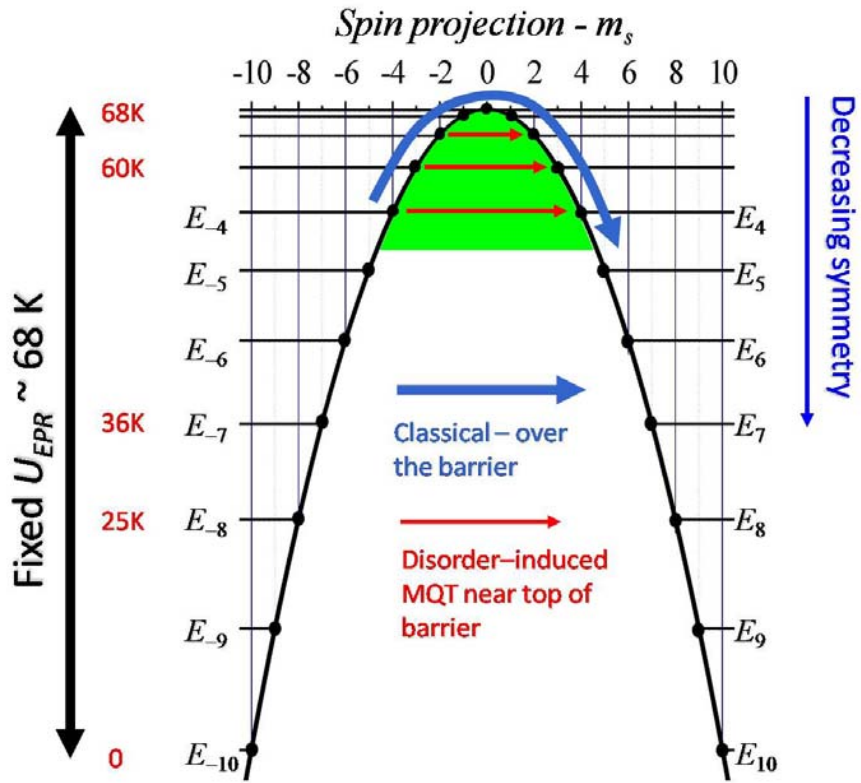

Figure 6, Redler et al.

17